\documentclass[11pt]{article}
\usepackage{graphicx}

\begin{document} 

\def\xslash#1{{\rlap{$#1$}/}}
\def \p {\partial}
\def \dd {\psi_{u\bar dg}}
\def \ddp {\psi_{u\bar dgg}}
\def \pq {\psi_{u\bar d\bar uu}}
\def \jpsi {J/\psi}
\def \psip {\psi^\prime}
\def \to {\rightarrow}
\def \lrto{\leftrightarrow} 
\def\bfsig{\mbox{\boldmath$\sigma$}}
\def\DT{\mbox{\boldmath$\Delta_T $}}
\def\xit{\mbox{\boldmath$\xi_\perp $}}
\def \jpsi {J/\psi}
\def\bfej{\mbox{\boldmath$\varepsilon$}}
\def \t {\tilde}
\def\epn {\varepsilon}
\def \up {\uparrow}
\def \dn {\downarrow}
\def \da {\dagger}
\def \pn3 {\phi_{u\bar d g}}

\def \p4n {\phi_{u\bar d gg}}

\def \bx {\bar x}
\def \by {\bar y}

\begin{center}
{\Large\bf  Gauge Invariance and QCD Twist-3 Factorization for Single Spin Asymmetries }
\par\vskip20pt
J.P. Ma$^{1,2}$ and G.P. Zhang$^{2}$     \\
{\small {\it
$^1$ State Key Laboratory of Theoretical Physics, Institute of Theoretical Physics, Academia Sinica,
P.O. Box 2735,
Beijing 100190, China\\
$^2$ Center for High-Energy Physics, Peking University, Beijing 100871, China  
}} \\
\end{center}
\vskip 1cm
\begin{abstract}
The collinear factorization at twist-3 for Drell-Yan processes is studied  with 
the motivation to solve the discrepancy in literature about the single spin asymmetry in the lepton angular distribution,  and 
to show how QCD gauge invariance is realized in the hadronic tensor.  The obtained result here agrees with 
our early result derived with a totally different approach.  In addition to the asymmetry we can construct another two observables to identify the spin effect. We show that the gauge invariance of different contributions in the hadronic tensor 
is made in different ways by summing the effects of gluon exchanges. More interestingly is that we can show that the virtual correction to one structure function of the 
hadronic tensor, hence to some weighted SSA observables, is completely determined by the quark form factor. This will simplify the calculation 
of higher order corrections. The corresponding result in semi-inclusive DIS is also given 
for the comparison with Drell-Yan processes.

\vskip 5mm
\end{abstract}
\vskip 1cm

\par\vskip20pt
In processes with a transversely polarized hadron in the initial state a Single Spin Asymmetry(SSA) can appear. 
This asymmetry is of particular interest in theory and experiment. Its existence requires nonzero absorptive part in scattering amplitudes and spin-flip interactions. For processes involving large scale, one can make theoretical predictions in terms of spin-dependent matrix elements in the framework of   
collinear factorization\cite{EFTE,QiuSt}. These matrix elements describe quark-gluon correlations 
in the transversely polarized hadron. Therefore, measuring SSA will reveal the correlations, hence the inner structure of hadrons.      

\par 
The collinear factorization of SSA is of twist-3. In general factorizations beyond twist-2 can be nontrivial even at the leading order of $\alpha_s$. This is partly reflected by the fact that different results of SSA 
in the lepton angular distribution of Drell-Yan processes exist in literature. In this work, we make a 
study of twist-3 factorization of SSA at the leading order of $\alpha_s$. The purpose is not only to solve 
the discrepancy but also to explicitly show gauge invariance of QCD and QED in the relevant hadronic tensor. We will show that the QCD gauge invariance is realized in different ways for different contributions to SSA. 
The obtained hadronic tensor for SSA is explicitly gauge invariant of QED. The importance of showing gauge 
invariance of QCD and QED is not only for obtaining consistent results in theory, but also for simplifying  the calculation of QCD corrections at certain level as we will show. For comparison with results for Drell-Yan processes 
we also give corresponding results for SSA in Semi-Inclusive DIS(SIDIS). 

\par

We consider the Drell-Yan process:
\begin{equation}
  h_A ( P_A, s) + h_B(P_B) \to \gamma^* (q) +X \to  \ell^-(k_1)  + \ell ^+(q-k_1)  + X,
\end{equation}
where $h_A$ is a spin-1/2 hadron with the spin-vector $s$.
To study the process it is convenient to use the  light-cone coordinate system, in which a
vector $a^\mu$ is expressed as $a^\mu = (a^+, a^-, \vec a_\perp) =
((a^0+a^3)/\sqrt{2}, (a^0-a^3)/\sqrt{2}, a^1, a^2)$ and $a_\perp^2
=(a^1)^2+(a^2)^2$. In this system we introduce two light-cone vectors 
$l^\mu =(1,0,0,0)$ and $n^\mu = (0,1,0,0)$. With the two vectors one can define 
the metric $g_\perp^{\mu\nu}$ and the totally anti-symmetric tensor $\epsilon_\perp^{\mu\nu}$ in the transverse 
space: 
\begin{equation}
  g_\perp^{\mu\nu} = g^{\mu\nu} - n^\mu l^\nu - n^\nu l^\mu,
  \quad 
  \epsilon_\perp^{\mu\nu} =\epsilon^{\alpha\beta\mu\nu}l_\alpha n_\beta, \quad \epsilon_\perp^{12}=-\epsilon_\perp^{21} =1. 
\end{equation}
We take a frame in which the momenta of hadrons and the spin $s$ are given by: 
\begin{equation}
  P_A^\mu \approx (P_A^+,0,0,0), \quad P_B^\mu  \approx (0,P_B^-,0,0), \quad s^\mu = s_\perp^\mu = (0,0, s_\perp^1, s_\perp^2). 
\end{equation}
We will look at the angular distribution in Collins-Soper frame\cite{CS-frame}. The solid angle of the lepton 
is given by $\Omega=(\theta,\phi)$. For the angular distributions studied here, we take the $x$-axis as the direction 
of the transverse spin, i.e., $s_\perp^\mu =(0,0,s_\perp^1,0)$. The $3$-momentum of the outgoing lepton $\ell^-$ in Collins-Soper frame 
is given by $ Q/2 (\sin\theta\cos\phi,\sin\theta\sin\phi,\cos\theta)$. 
The differential angular distribution can be written as:
\begin{equation}
\frac{d\sigma}{d Q^2 d\Omega} = \frac{\alpha^2  e_q^2}{4 S Q^4} \int d^4  q \delta (q^2-Q^2)  W^{\mu\nu} L_{\mu\nu}
\label{ADis} 
\end{equation} 
where $L_{\mu\nu}$ is the leptonic tensor $2 ( 2 k_1^\mu (q-k_1)^\nu + 2 k_1^\nu (q-k_1)^\mu -  g^{\mu\nu} Q^2)$. $e_q$ is electric charge fraction of the quark.  $S$ is the invariant mass $S=(P_A+P_B)^2$.
 $W^{\mu\nu}$ is the hadronic tensor defined as 
\begin{eqnarray}
  W^{\mu\nu} =  \sum_X \int \frac{d^4 x}{(2\pi)^4} e^{iq \cdot x} \langle h_A (P_A, s_\perp), h_B(P_B)  \vert
    \bar \psi(0) \gamma^\nu \psi(0) \vert X\rangle \langle X \vert \bar \psi (x) \gamma^\mu \psi(x) \vert
     h_B(P_B),h_A (P_A, s_\perp)  \rangle.  
\end{eqnarray}
The invariant mass of the lepton pair is $Q^2$. The SSA in the 
angular distribution is defined as: 
\begin{equation} 
A_N = \left (\frac{d\sigma( \vec s_\perp) }{d Q^2 d\Omega} - \frac{d\sigma( -\vec s_\perp) }{d Q^2 d\Omega} \right )
   \biggr / 
 \left ( \frac{d\sigma( \vec s_\perp) }{d Q^2 d\Omega} + \frac{d\sigma( -\vec s_\perp) }{d Q^2 d\Omega} \right ).
\label{AN} 
\end{equation} 
In the collinear factorization, the spin-dependent part of the differential cross-section at the leading order 
of $\alpha_s$ can be factorized 
with the ETQS matrix elements\cite{EFTE,QiuSt}. They are defined with QCD operators of twist-3. Not all ETQS matrix elements 
are independent, among them there are relations\cite{EKT,ZYL}.  One of the 
matrix elements is relevant in this work. It is defined as: 
\begin{eqnarray}
T_F (x_1,x_2) \tilde s_\perp^\mu  
   =    g_s  \int \frac{d\lambda_1 d\lambda_2}{4\pi}
   e^{ -i\lambda_2 (x_2-x_1) P^+_A -i \lambda_1 x_1 P^+_A }
   \langle P_A, \vec s_\perp \vert
           \bar\psi (\lambda_1n ) \gamma^+ G^{+\mu}(\lambda_2n) \psi(0) \vert P_A,\vec s_\perp \rangle 
\label{ETQS}
\end{eqnarray} 
with $\tilde s_\perp^\mu = \epsilon_\perp^{\mu\nu} s_{\perp\nu}$. In the definition we have suppressed 
the gauge links. In the light-cone gauge $n\cdot G$=0 they are units. In other gauges one needs to insert 
the product of gauge links like
\begin{equation} 
    {\mathcal L}_n (x) = P\exp\left \{ -i g_s \int_{-\infty}^0 d\lambda n\cdot G(\lambda n +x) \right \} 
\label{GL}                                
\end{equation} 
between operators to make the definition gauge invariant. In this work we will work in Feynman gauge.

\par 
The asymmetry $A_N$ has been studied in several works\cite{DY1,DY2,DY3,DY4,AT,ZM,MaZh,CR}. As mentioned, different results exist. For simplicity 
we will only consider the case where the process is initiated by an anti-quark from $h_B$. 
The different results can be given as
\begin{equation} 
\frac{1}{2} A_N^{\cite{AT,CR}} = A_N^{\cite{DY2,DY3}}
  = 2 A_N^{\cite{ZM,MaZh}} = -\frac{\sin(2\theta)\sin\phi } {Q(1+\cos^2\theta)} 
    \frac{ \displaystyle{  \int d x d y T_F(x,x) \bar q(y) \delta ( xyS-Q^2) } }
      {  \displaystyle { \int dx dy q(x) \bar q(y) \delta ( xyS-Q^2) } }. 
\label{BQ} 
\end{equation} 
where $\bar q(y)$ is the anti-quark distribution of $h_B$ and the numbers in $[\cdots]$'s are reference numbers. 
In \cite{DY1} there is an additional contribution with the derivative of $T_F(x,x)$. It is noted that 
in the works except ours in \cite{MaZh} $A_N$ is derived with the diagram expansion at the hadron level. In \cite{MaZh} 
a different approach is taken by replacing $h_A$ with a multi-parton state and $h_B$ with an anti-quark. 
Then one can calculate $A_N$ and $T_F$ with these parton states separately to find the factorized form.
Since the hadron $h_A$ is transversely polarized, one will get null results for $A_N$ and $T_F$ 
if one replaces $h_A$ with a single quark because of the helicity conservation of QCD. To obtain nonzero results, one has to use  a multi-parton state instead of the single quark\cite{CMS}. This approach is useful to clarify some discrepancies in results about SSA, e.g., 
in the evolution of $T_F(x,x)$\cite{MW}.        
\par 
\begin{figure}[hbt]
\begin{center}
\includegraphics[width=14cm]{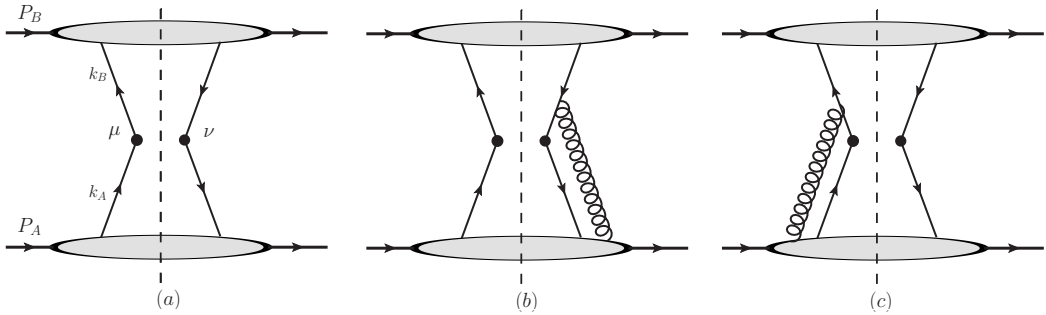}
\end{center}
\caption{The tree-level diagrams for $W^{\mu\nu}$ of Drell-Yan processes. The black dots represent the insertion of the electromagnetic current.   } 
\label{Feynman-dg1}
\end{figure}
 
\par 
In this work we will use the diagram expansion at hadron level to derive the hadronic tensor and $A_N$. 
At the leading order of $\alpha_s$ one has the contributions to $W^{\mu\nu}$ from diagrams given by Fig.1. 
These diagrams can be divided into three parts: the upper- and lower bubble, and the middle part given by 
explicit Feynman diagrams of parton scattering.  
The upper- and lower bubble represent jet-like Green functions related 
to the initial hadron $h_B$ and $h_A$, respectively. By imaging hadrons as bound states of quarks and gluons, 
these bubbles are the sum of all possible diagrams involving corresponding hadrons. Since they are jet-like, 
there are power counting for momenta of partons leaving or entering the bubbles. E.g., in Fig.1a the momenta 
$k_A$ and $k_B$ scale like: 
\begin{equation} 
 k_A^\mu  \sim Q (1,\lambda^2,\lambda,\lambda), \quad  k_B^\mu \sim Q (\lambda^2, 1,\lambda,\lambda),
\label{PC} 
\end{equation}
with $\lambda \sim \Lambda_{QCD}/Q$. For Fig.1b or 1c, where a gluon leaves or enters bubbles, the gluon 
field vectors also scale like the pattern of the corresponding momentum as in Eq.(\ref{PC}) in the 
gauge we work. In collinear factorization one needs to expand the contributions from Fig.1 
in power of $\lambda$. We will call the gluons with the polarization index $-$ or $+$ as $G^-$- and $G^+$-gluons.            
With the power counting one easily finds that the contributions from exchange of any number of 
$G^-$ gluons with the upper bubble or $G^+$-gluons with the lower bubble can be at the same power of $\lambda$. The summation
of the contributions 
is needed. In the first step, we only consider diagrams in Fig.1.     
\par 
In our case up to twist-3, we can always make the approximation by taking the twist-2 part of the upper bubble 
of $h_B$ and neglecting all components of $k_B$ except the $-$-component. The twist-2 part is given by the 
anti-quark distribution function $\bar q (y)$ of $h_B$ with $k_B^-=y P_B^-$. For momenta of partons from the lower bubble of $h_A$, 
one can always neglect the $-$-components. With this in mind the contribution from Fig.1a 
can be written as:
\begin{eqnarray} 
 W^{\mu\nu}\biggr\vert_{1a} = \int d^3 \tilde k_A  d k_B^- \left ( \frac{1}{2 N_c} \bar q(y) \right )  \biggr [  \delta^4 (k_A +k_B -q) 
 \left ( \gamma^\nu  \gamma^+ \gamma^\mu \right )_{ji}   \biggr ]  
 \biggr [ \int \frac{d^3 \tilde \xi}{(2\pi)^3}  e^{-i\xi\cdot k_A} \langle h_A\vert \bar \psi_j(\xi) \psi_i(0) \vert h_A  \rangle \biggr ] , 
\label{1a} 
\end{eqnarray}       
with the notations:
\begin{equation}
   k_B^\mu =(0,k_B^-,0,0),\quad k_A^\mu =(k_A^+,0, \vec k_{A\perp}), \quad 
   \xi^\mu = (0,\xi^-,\vec \xi_\perp), \quad d^3\tilde \xi = d\xi^- d^2\xi_\perp, \quad d^3\tilde k_A 
    = d k_A^+ d^2 k_{A\perp}. 
\end{equation}
$ij$ are the indices of Dirac spinor and color.     
In Eq.(\ref{1a}) the term in $(\cdots)$ combined with $\gamma^+$ belongs to the twist-2 part of the upper 
bubble,  the term in the first $[\cdots ]$ without $\gamma^+$ are from the middle part of Fig.1a. 
The term in the second $[\cdots ]$ is from the lower bubble. 
In the expansion of $\lambda$ one expands not only the middle part but also the bubbles in $\lambda$. 
The upper bubble is already expanded and the leading order is taken. The leading order of the lower bubble 
is given by taking the matrix as $\gamma^-$ and keeping the $+$-component of $k_A$ as nonzero.   
One then obtains the standard result for 
the spin-independent part of $W^{\mu\nu}$ at twist-2:
\begin{eqnarray} 
 W^{\mu\nu}\biggr\vert_{twist-2} = - \frac{1}{N_c} g_\perp^{\mu\nu} \delta^2 (q_\perp)  q(x) \bar q(y), 
\quad   q^+ = xP_A^+, \quad q^- =y P_B^-. 
\label{t2} 
\end{eqnarray}   
It is noted that in this result the sum of contributions from exchange of any number of $G^+$-gluons with the 
lower bubble and $G^-$-gluons with the upper bubble has been performed. It results in that there are gauge links 
in the definition of the parton distribution functions $q(x)$ and $\bar q(y)$\cite{PDFFF}. 

\par 
It is clear that Fig.1a also contains contributions at the next-to-leading order of $\lambda$ or twist-3, if we take 
the middle part or the lower bubble at the Next-to-Leading Order(NLO) of $\lambda$.  To obtain the twist-3 contribution it is convenient to write the product of $\Gamma$-matrices in Eq.(\ref{1a}): 
\begin{eqnarray} 
 \gamma^\nu \gamma^+ \gamma^\mu = -\gamma^+ g_\perp^{\mu\nu} + n^\mu \gamma_\perp^\nu + n^\nu \gamma_\perp^\mu 
    + 2 n^\mu n^\nu \gamma^- -i\epsilon^{\alpha \rho \mu\nu} \gamma_5 \gamma_\rho n_\alpha. 
\label{GD}    
\end{eqnarray}
In the above only the first three terms can give contributions at NLO of $\lambda$. For the contribution 
from the first term one needs to expand the middle part in $\lambda$. The NLO term in the expansion comes 
from the expansion of the $\delta$-function. 
We denote this contribution with the index $1a^+$, which is given as: 
\begin{eqnarray} 
 W^{\mu\nu}\biggr\vert_{1a^+}  = \frac{i}{2 N_c}  \bar q(y) 
 \frac{\partial \delta^2 (q_\perp)}{\partial q_\perp^\rho} g_\perp^{\mu\nu}    
 \int \frac{d \lambda }{2\pi}  e^{ i\lambda k_A^+ } \langle h_A\vert   \bar \psi  (0 ) \gamma^+ \partial_\perp^\rho \psi(\lambda n ) \vert h_A \rangle .
\label{1a+} 
\end{eqnarray}
The contribution from the second- and third term in Eq.(\ref{GD}) involves the matrix element of the operator $\bar \psi(\lambda n)\gamma_\perp^\mu \psi (0)$.  As in the case of twist-2 contribution, the exchange 
of $G^+$-gluons from the lower bubble will contribute at the same power of $\lambda$. These contributions need  
be summed with gauge links. After the summation, the contribution from $\gamma_\perp$ in Eq.(\ref{GD}) involves the matrix element of $\bar \psi(\lambda n) {\mathcal L}_n (\lambda n) \gamma_\perp^\mu  {\mathcal L}_n^\dagger (0) \psi (0)$.  
Since the $\gamma$-matrix here is $\gamma_\perp$, the contributions from exchanges of $G^+$-gluons is not included in the twist-3 contributions from Fig.1b and Fig.1c, where the involved $\gamma$-matrix is $\gamma^+$ as shown later. 
\par  
To find the twist-3 contributions from $\gamma_\perp$ in Eq.(\ref{GD}) one needs to separate quark fields into 
the good- and bad component\cite{JiOs}:  
\begin{equation} 
   \psi =\psi_+ +\psi_-, \quad \psi_+ =\frac{1}{2} \gamma^- \gamma^+ \psi, \quad   \psi_- =\frac{1}{2} \gamma^+ \gamma^- \psi, 
\end{equation}
where $\psi_+$ is the good component and $\psi_-$ is the bad component. 
Using the equation of motion one can solve $\psi_-$ in terms of $\psi_+$: 
\begin{equation} 
   \psi_-(\xi) = -\frac{1}{2} {\mathcal L}_n(\xi) \int_{-\infty}^0 d\lambda {\mathcal L}_n^\dagger (\lambda n +\xi) 
     \gamma^+ \gamma_\perp\cdot D_\perp \psi_+ (\lambda n + \xi). 
\end{equation} 
With the solution the contribution from $\gamma_\perp$ is proportional to: 
\begin{eqnarray} 
&& \int \frac{ d \xi^-} {2\pi} e^{-i \xi^- k_A^+}  \langle h_A \vert \bar \psi(\xi^- n) {\mathcal L}_n (\xi^- n)  \gamma_\perp^\mu  {\mathcal L}_n^\dagger (0) \psi(0) \vert h_A \rangle  
\nonumber\\
 && =- \frac{i}{2 k_A^+} \int\frac{d\xi^-} {2\pi} e^{-i\xi^-k_A^+} 
     \biggr [ \langle h_A \vert \bar \psi(\xi^- n) {\mathcal L}_n(\xi^-n) \gamma_\perp^\mu \gamma^+ {\mathcal L}_n^\dagger (0) \gamma_\perp\cdot D_\perp \psi (0) \vert h_A \rangle 
\nonumber\\
 &&  -    \langle h_A \vert \overline{( \gamma_\perp \cdot D_\perp \psi)} (\xi^- n) {\mathcal L}_n(\xi^-n) 
     \gamma^+  \gamma_\perp^\mu {\mathcal L}_n^\dagger (0)  \psi (0) \vert h_A \rangle \biggr ] . 
\label{GAT} 
\end{eqnarray}
Using symmetries of Parity(P) and Time-reversal(T), one can show that the combination of matrix elements
in $ [\cdots ]$ does not depend on the transverse spin $s_\perp$. The combination is zero. 
In the product $ {\mathcal L}_n (\xi^- n) {\mathcal L}_n^\dagger (0)$ the direction of the gauge link ${\mathcal L}_n$ is pointing to the past as shown in Eq.(\ref{GL}).   
It is noted that in the collinear factorization 
with non-singular gauge like Feynman gauge, it is equivalent by taking the direction of gauge links pointing 
to the future in the product. The reason for this is that one can use the products of the gauge links to form a gauge link 
along a closed contour and the enclosed area is zero. In fact, in non-singular gauges one can show with PT-symmetry the contribution in Eq.(\ref{1a+}) is also zero. Therefore, we conclude that 
the total contribution from Fig.1a is zero in non-singular gauges.       
\par

\par 
The contributions from Fig.1b and Fig.1c after the approximation made similarly for Eq.(\ref{1a}) are:
\begin{eqnarray}
W^{\mu\nu}\biggr\vert_{1c} &=& \int d^3 \tilde  k_A d^3 \tilde k d k_B^-   \frac{1}{2 N_c} \bar q(y) 
   \biggr   [ \delta^4 (  k_A +   k +  k_B -q) \gamma^\nu \gamma^+ 
          \gamma_\rho   \frac{\gamma\cdot (-  k - k_B)}{( k+  k_B)^2 + i\varepsilon }\gamma^\mu \biggr ]_{ji}
          {\mathcal M}_{ij}^\rho (  k_A,  k), 
\nonumber\\
W^{\mu\nu}\biggr\vert_{1b} &=& \int d^3 \tilde  k_A d^3 \tilde k d k_B^-  \frac{1}{2 N_c} \bar q(y) 
     \biggr [ \delta^4 ( k_A +  k_B -q) \gamma^\nu 
            \frac{\gamma\cdot ( k - k_B)}{( k -  k_B)^2 - i\varepsilon }\gamma_\rho \gamma^+  \gamma^\mu \biggr ]_{ji}{\mathcal M}_{ij} ^\rho ( k_A,  k) ,  
\label{1c1b}                    
\end{eqnarray} 
with the quark-gluon correlator  
\begin{eqnarray}             
 {\mathcal M}_{ ij}^{\rho} (  k_A,  k) = g_s\int \frac{d\xi^-_1 d^2 \xi_{1\perp} d\xi_2^- d^2\xi_{2\perp} }{(2\pi)^6} 
     e^{ i  \xi_1 \cdot  k_A +i \xi_2 \cdot  k } 
       \langle h_A \vert \bar \psi_j (0 ) \left [  G^{\rho} (\xi_2 ) \psi ( \xi_1 ) \right ] _i \vert h_A \rangle \biggr\vert_{\xi_{1,2}^+=0},         
\end{eqnarray}
where $k$ is the momentum of the gluon. It is given by $k^\mu =(k^+,0,\vec k_\perp)$ and  the measure $d^3\tilde k = d k^+ d^2k_\perp$. $ij$ are the indices of Dirac spinor and color. In Eq.(\ref{1c1b}) 
the terms in $[\cdots ]$ are from the middle part of diagrams, and ${\mathcal M}^\rho$ is for the lower bubble. 
To find the relevant contributions one can represent the matrix ${\mathcal M}^{\rho}$ as: 
\begin{eqnarray}
{\mathcal M}^\rho ( k_A, k) = \frac{1}{2N_c} \biggr ( M^\rho (k_A, k)  \gamma^- 
  + M_A^\rho( k_A,  k)\gamma_5 \gamma^- \biggr )   
  +\cdots, 
\end{eqnarray}
the $\cdots$ represent terms which will give contributions beyond twist-3. The term with the matrix 
of $\gamma_5 \gamma^-$ gives no contribution at leading order of $\alpha_s$. Here the involved operator 
is with $\gamma^+$ as mentioned after Eq.(\ref{1a+}).    
If we expand the contributions in Eq.(\ref{1c1b}) in $\lambda$, the contributions at the leading order 
or at twist-2 are spin-independent and only contain $M^+$. These contributions are summed in Eq.(\ref{t2}) as gauge links in $q(x)$.  
\par 
At NLO of $\lambda$ or at twist-3 the contributions can be spin-dependent.  
To obtain the contributions at twist-3 one needs to expand the middle parts in Eq.(\ref{1c1b}) 
at NLO and the lower bubble ${\mathcal M}^\rho$ at NLO. One should note that in the expansion of the middle 
part the $\delta$-function depends on $k_{A\perp}$ and $k_\perp$. This dependence also needs to be 
expanded. With the power counting for the gluon field explained after Eq.(\ref{PC})  $M^\rho$ has the NLO 
contribution with $\rho=\perp$. Keeping these in mind the expansion is straightforward. We obtain 
the contributions from Fig.1b and Fig.1c as: 
\begin{eqnarray} 
W^{\mu\nu}\biggr\vert_{1c} &=& \frac{1 }{ N_c } \bar q(y) \int d k^+ \left (  \frac{i} {k^+  + i\varepsilon } \right )   
 \biggr \{  i\frac{\partial \delta^2 (q_\perp)}{\partial q_\perp^\rho}g_\perp^{\mu\nu}  \int d^2  k_{A\perp} d^2  k_\perp   
  \left ( k_{A\perp}^\rho +k_\perp^{\rho} \right )
     M^+(k_A-k^+l ,k)   
\nonumber\\ 
  &&+\delta^2 (q_\perp) \frac{ l^\mu}{k_B^-}  \biggr [  g_s \int\frac{d \xi_1^- d\xi_2^-} {2 (2\pi)^2}  
     e^{ i  \xi_1^- (k_A^+-k^+) +i\xi_2^- k^+ } 
       \langle h_A \vert \bar \psi (0 ) \gamma^+  \hat G^{+\nu} (\xi_2^- n)  \psi ( \xi_1^- n  ) \vert h_A \rangle \biggr ]         
      \biggr\}  +\cdots,
\nonumber\\
W^{\mu\nu}\biggr\vert_{1b} &=& \frac{1 }{ N_c  } \bar q(y) \int d k^+ \left (  \frac{-i} {-k^+  - i\varepsilon } \right )    
 \biggr \{  -i\frac{\partial \delta^2 (q_\perp)}{\partial q_\perp^\rho} g_\perp^{\mu\nu}\int d^2  k_{A\perp} d^2  k_\perp   
 k_{A\perp}^\rho 
    M^+(k_A, k)  
\nonumber\\ 
   && +\delta^2 (q_\perp)  \frac{l^\nu}{k_B^-} \biggr [ g_s \int\frac{d \xi_1^- d\xi_2^-} {2 (2\pi)^2}  
     e^{ i  \xi_1^- k_A^+ +i\xi_2^- k^+ } 
       \langle h_A \vert \bar \psi (0 ) \gamma^+  \hat G^{+\mu} (\xi_2^- n)  \psi ( \xi_1^- n  ) \vert h_A \rangle \biggr ] 
 \biggr \}     
  + \cdots, 
\label{1c1bexp}                                        
\end{eqnarray}
with 
\begin{equation} 
  \hat G^{\mu\nu} =\partial^\mu G^\nu -\partial^\nu G^\mu, \quad k_A^+ =q^+= x P_A^+, \quad k_B^-=q^-= y P_B^-.  
\end{equation} 
In Eq.(\ref{1c1bexp}) $\cdots$ stand for higher order of $\lambda$ or spin-independent parts. 
\par 
The results in Eq.({\ref{1c1bexp}) are not expressed with matrix elements of gauge invariant operators.  
E.g., in the second- and fourth line one has the operator $\hat G^{\mu\nu}$ which is not exactly the gluon field strength tensor $G^{\mu\nu}$. It is commonly believed that one will have the results
expressed with gauge invariant operators after summing of any number 
of gluon exchanges additionally to the exchange of one gluon in Fig.1b and Fig.1c. This gives a common rule 
for obtaining gauge invariant results:  One only calculates Fig.1b and Fig.1c with the exchange of a $G^+$-gluon in Feynman gauge and obtains the combination at NLO of $\lambda$ $k_\perp^\mu G^+$. Then one simply makes the replacement $k_\perp^\mu G^+ \to G^{+\mu}$ in the results, e.g., in \cite{DY3,ZM}. Although the replacement 
will produce correct results here, but this rule is not fully augmented to be correct. This brings up the question if the results are really QCD gauge invariant. 
\par 
Taking the contributions of Fig.1c as an example, one indeed obtains for the contribution with $\delta^2(q_\perp)$ the full gluon field strength tensor $G^{+\mu}$ instead of $\hat G^{+\mu}$ and gauge links between operators,
after the summation of any number of gluon exchanges. After taking the cut of the quark propagator 
in the middle part of Fig.1c, one obtains the contribution expressed with $T_F(x,x)$. 
But for the contribution with the derivative of $\delta^2(q_\perp)$ the exchanged gluons are all $G^+$-gluons.  The contribution from the exchange of one or 
more gluons with the transverse polarization is beyond the order of $\lambda$ considered here.  
Therefore, one can not use the 
common rule in a direct way to obtain a gauge invariant result from this contribution. In fact 
the contribution from the exchange of $G^+$-gluons can be easily summed with gauge links. The summation 
is like that in the factorization at twist-2. Therefore, after the summation of $n$-gluon exchanges with 
$n>0$, we have the sum of Fig.1b and Fig.1c: 
\begin{eqnarray}  
W^{\mu\nu}\biggr\vert_{1b+1c } &=& \frac{1 }{2 N_c k_B^- } \delta^2 (q_\perp) 
\left ( l^\mu \tilde s^\nu_\perp + l^\nu \tilde s_\perp^\mu \right ) \bar q(y) T_F (x,x)
  + \frac{i }{ N_c } g_\perp^{\mu\nu} \bar q (y)\frac{\partial \delta^2 (q_\perp)}{\partial q_\perp^\rho}  \int \frac{d\lambda} { 4\pi} e^{  i k_A^+ \lambda } 
\nonumber\\       
  &&  
 \biggr [  \langle h_A \vert \bar \psi (0)  {\mathcal L}_n (0) 
  \gamma^+ 
     \partial_\perp^\rho  \left ({\mathcal L}_n^\dagger  (\lambda n )       \psi ( \lambda n ) \right )
          \vert h_A \rangle
 - \langle h_A \vert \bar \psi (0)  
 \gamma^+ 
     \partial_\perp^\rho   \psi ( \lambda n ) 
          \vert h_A \rangle \biggr ].  
\label{1b+1c}                        
\end{eqnarray} 
In a non-singular gauge which we use the first term in $[\cdots]$ is gauge invariant. 
The second term  in $[\cdots]$ is not gauge invariant, but it will be canceled in the sum with the contribution 
in Eq.(\ref{1a+}). In non-singular gauges it is zero.  
\par
To simplify our result in the above, we note that one can derive the identity in non-singular gauges: 
\begin{equation} 
 \partial_\perp^\mu \left ( {\mathcal L}_n^\dagger \psi \right ) (\xi) = {\mathcal L}_n^\dagger D_\perp^\mu \psi (\xi) - i g_s \int_{-\infty}^0 d\lambda \left ( {\mathcal L}_n^\dagger G^{+\mu} {\mathcal L}_n \right ) 
   (\lambda n + \xi)  \left ( {\mathcal L}_n^\dagger \psi \right ) (\xi).
\label{IDEN6}
\end{equation} 
Using this identity and PT-symmetry we can derive our final result of the twist-3 contribution of $W^{\mu\nu}$ from Fig.1
as:
\begin{eqnarray}  
W^{\mu\nu}\biggr\vert_{Fig.1}  = \frac{1 }{2 N_c}\bar q(y)  T_F (x,x)  \biggr [ \frac{1}{ q^- } \delta^2 (q_\perp) 
\left ( l^\mu \tilde s^\nu_\perp + l^\nu \tilde s_\perp^\mu \right ) 
  +  g_\perp^{\mu\nu} \frac{\partial \delta^2 (q_\perp)}{\partial q_\perp^\rho} 
    \tilde s_\perp^\rho \biggr ] . 
\label{FW}          
\end{eqnarray}
This result is QCD gauge invariant. 
From our derivation one can see that the way to obtain final results with matrix elements of QCD gauge invariant operators is different for different contributions, although with the common rule mentioned before one obtains 
the same results. In the summation of exchanges of gluons for the contribution with $\delta^2 (q_\perp)$, 
one of the exchanged gluons is transversely polarized. This enables us immediately to obtain the matrix element 
of the gauge invariant operator with $G^{+\mu}$. In the summation of exchanges of gluons for the contribution with the derivative of $\delta^2 (q_\perp)$, all exchanged gluons are $G^+$-gluons, the summed form does not involve $G^{+\mu}$ as shown in Eq.(\ref{1b+1c}). Only after using the identity and PT-symmetry, one can express 
the contribution with $T_F(x,x)$. This has a consequence  which we will discuss after that we have given the relevant results for SIDIS.  
\par 
In the contribution of Fig.1, $T_F(x,x)$ is defined with the chirality-even operator. There is also a twist-3 contribution involving matrix elements defined with chirality-odd operators\cite{ZM}. One of the involved distributions defined with chirality-odd operators is the transversity distribution of $h_A$: 
\begin{eqnarray} 
 h_1 (x) s_\perp^\mu = \int \frac{d \xi^-}{4\pi}  e^{-ix \xi^- P_A^+ } \langle h_A\vert \bar q(\xi^- n)\gamma_5 
 \gamma^+\gamma_\perp^\mu q(0) \vert h_A \rangle.  
\end{eqnarray}
Another one is the chirality-odd quark-gluon correlation of $h_B$ defined as:  
\begin{equation}  
  T_{F}^{(\sigma) } (y_1,y_2) = -g_s \int\frac{ d\xi_1^+ d\xi_2^+}{4\pi} e^{-i \xi_1^+ y_1 P_B^- -i \xi_2^+(y_2-y_1) 
     P_B^- }
     \langle h_B\vert \bar q(0) \left (i\gamma_{\perp\mu} \gamma^- \right ) G^{+\mu}(\xi_2^+ l)  q(\xi_1^+ l) \vert h_B \rangle.
\label{COTF}           
\end{equation}
In the above we have suppressed gauge links. 
The chirality-odd contribution to $W^{\mu\nu}$ can be derived in a similar way. Adding this contribution we have the complete result for the spin-dependent part of $W^{\mu\nu}$: 
\begin{eqnarray}
W^{\mu\nu}  &=& \frac{1 }{2 N_c} \biggr \{ - T_F^{(\sigma)} (y,y) h_1(x) \biggr [ \frac{1}{2} \frac{\partial \delta^2 (q_\perp)}{\partial q_\perp^\rho } \biggr ( g_\perp^{\mu\rho} \tilde s_\perp^\nu +g_\perp^{\nu\rho}\tilde s_\perp^\mu 
     -g_\perp^{\mu\nu} \tilde s_\perp^\rho \biggr ) +  \frac{\delta^2 (q_\perp)}{P_B\cdot q} \biggr ( P_B^\mu \tilde s_\perp^\nu + P_B^\nu\tilde s_\perp^\mu \biggr ) \biggr ]  
\nonumber\\
   && + \bar q(y)  T_F (x,x)  \biggr [ \frac{\delta^2 (q_\perp)}{ P_A\cdot q  }  
\biggr ( P_A^\mu \tilde s^\nu_\perp + P_A^\nu \tilde s_\perp^\mu \biggr ) 
  +  g_\perp^{\mu\nu} \frac{\partial \delta^2 (q_\perp)}{\partial q_\perp^\rho} 
    \tilde s_\perp^\rho \biggr ]\biggr \} + {\mathcal O}(\alpha_s). 
\label{TOW}          
\end{eqnarray}  
With the explicit result for $W^{\mu\nu}$ one can check if it is $U(1)$-gauge invariant. However, 
one can not simply contract $q_\mu$ with $W^{\mu\nu}$ here, because $W^{\mu\nu}$ contains $\delta^2(q_\perp)$ 
and its derivative. One should take $W^{\mu\nu}$ as a distribution tensor. The  $U(1)$-gauge invariance should be understood in the sense of integration of $q_\perp$. Therefore, the  $U(1)$-gauge invariance implies: 
\begin{equation} 
  \int d^2 q_\perp  {\mathcal F}(q_\perp)W^{\mu\nu} q_{\mu} =0. 
\label{U1}  
\end{equation}
with ${\mathcal F}(q_\perp)$ as a test function.      
Our result satisfies this equation. 
\par 
From our result of $W^{\mu\nu}$ we can calculate the asymmetry $A_N$. The result is with $s_\perp^\mu =(0,0,1,0)$:
\begin{equation} 
A_N = -\frac{\sin(2\theta)\sin\phi } {2 Q(1+\cos^2\theta)} 
    \frac{ \displaystyle{  \int d x d y \left [ T_F(x,x) \bar q(y) +h_1 (x) T_F^{(\sigma)} (y,y) \right ]  \delta ( xyS-Q^2) } }
      {  \displaystyle { \int dx dy q(x) \bar q(y) \delta ( xyS-Q^2) } }. 
\label{AN} 
\end{equation} 
This result agrees with that given in \cite{ZM}. Especially, the chiral-even contribution agrees with our 
early result in \cite{MaZh}. Since the same contribution has been derived in a completely different way, 
we believe that the discrepancy illustrated in Eq.(\ref{BQ}) is solved with Eq.(\ref{AN}). 
\par 
From our result $W^{\mu\nu}$ consists of different tensor structures at leading order of $\alpha_s$. 
This enables us to construct more observables. In the differential cross-section 
defined in Eq.(\ref{ADis}) $q_\perp$ is integrated. In the integration one can introduce 
some weight functions to have the so-called weighted SSA. 
In the case here, we can introduce two weighted SSA's. We derive: 
\begin{eqnarray} 
\frac{ d\sigma\langle q_\perp \cdot \tilde s_\perp \rangle }{ d Q^2 d \Omega} 
   &=&  \frac{\alpha^2 e_q^2}{4 S Q^4} \int d^4  q \delta (q^2-Q^2) \left ( q_\perp\cdot \tilde s_\perp \right ) 
 W^{\mu\nu} L_{\mu\nu}   
\nonumber\\
   &=& -\frac{\alpha^2 e_q^2}{8 N_c Q^2 } \int dx dy \delta (xy S-Q^2) \biggr [ 
     (1+\cos^2\theta ) \bar q(y) T_F (x,x) -\frac{1}{2} \sin^2\theta \cos(2\phi) h_1(x) T_F^{(\sigma)} (y,y)\biggr], 
\nonumber\\
\frac{ d\sigma\langle q_\perp \cdot s_\perp \rangle }{ d Q^2 d \Omega} 
   &=&  \frac{\alpha^2 e_q^2}{4 S Q^4} \int d^4  q \delta (q^2-Q^2) \left ( q_\perp\cdot s_\perp \right ) 
 W^{\mu\nu} L_{\mu\nu}  \nonumber\\
   &=&  -\frac{\alpha^2 e_q^2}{16 N_c Q^2 } \sin^2\theta \sin(2\phi) \int dx dy \delta (xy S-Q^2) h_1(x) T_F^{(\sigma)}(y,y). 
\end{eqnarray}    
Measuring these observables and $A_N$ will help to disentangle the chiral-even- and chiral-odd contribution. 
\par
\par
\begin{figure}[hbt]
\begin{center}
\includegraphics[width=14cm]{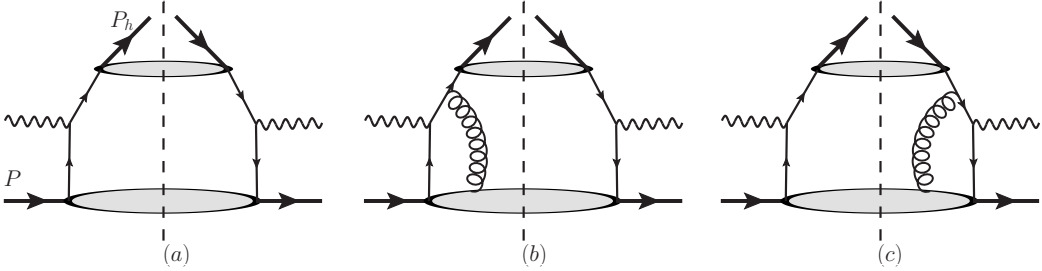}
\end{center}
\caption{Diagrams for contributions in SIDIS. } 
\end{figure}
\par

One can expect that there is a close relation between Drell-Yan process and SIDIS. To compare SSA in the two processes, we discuss here SSA in SIDIS briefly. 
We study here SSA in Semi-Inclusive DIS(SIDIS):
\begin{equation} 
   \ell (k) + h_A(P,s) \to \ell (k') + h_B (P_h) + X
\label{SIDIS}    
\end{equation} 
where the initial hadron is polarized with the spin vector $s$. We take the initial electron as unpolarized. 
The polarization of particles in the final state is not observed or 
summed. The standard variables for SIDIS are defined as:
\begin{equation} 
x_B = \frac{Q^2}{2 P\cdot q},\ \ \  y=\frac{ P\cdot q}{P\cdot k}, \ \ \ z_h=\frac{P\cdot P_h}{P\cdot q}.\end{equation} 
The hadronic tensor of SIDIS is defined as 
\begin{equation} 
W^{\mu\nu} = \sum_X \int \frac {d^4 x}{(2\pi)^4} e^{iq\cdot x} \langle h_A \vert J^\mu (x) \vert h_B, X\rangle 
     \langle X, h_B\vert J^\nu (0) \vert h_A \rangle. 
\end{equation} 
At leading order of $\alpha_s$, the spin-dependent part of $W^{\mu\nu}$ receives contributions from diagrams given in Fig.2.  
To analyze SIDIS, it is convenient to take the frame in which the initial hadron moves in the $z$-direction
and the final hadron moves in the $-z$-direction. The transverse space is  then defined with $P$ and $P_h$. In this frame the virtual photon has a transverse momentum. The result obtained in this frame can be expressed 
in a covariant way.  The calculation of Fig.2 is similar to that of Fig.1 for Drell-Yan processes. 
We will skip any detail of the calculation and give the results in the following. 
\par 
From Fig.2 one can derive the spin-dependent part of $W^{\mu\nu}$ as: 
\begin{equation} 
 W^{\mu\nu} =  \frac{1}{z_h} d(z_h) T_F (x_B,x_B) \biggr [ \delta^2 (q_\perp) \frac{1}{P\cdot q} ( P^\mu \tilde s_\perp^\nu 
   + P^\nu \tilde s_\perp^\mu ) + g_{\perp h}^{\mu\nu} \tilde s_{\perp h}^\rho \frac{\partial \delta^2 (q_\perp)}{\partial q_\perp^\rho}  \biggr ]. 
\label{WSIDIS}   
\end{equation}
with 
\begin{equation} 
  g_{\perp h} ^{\mu\nu} = g^{\mu\nu} -\frac{1}{P\cdot P_h} \biggr ( P^\mu P_h^\nu + P^\nu P_h^\mu \biggr ), 
  \quad \tilde s^{\mu}_{\perp h} =\frac{1}{P\cdot P_h} \epsilon^{\alpha\beta \mu\nu} P_\alpha P_{h\beta} s_\nu, 
  \quad q_\perp^\mu = g_{\perp h}^{\mu\nu} q_\nu. 
\label{GE}     
\end{equation} 
$d(z_h)$ is the quark fragmentation function of $h_B$. It is defined:   
\begin{equation} 
  d (z) = z\int\frac{d y}{4\pi} e^{- i y P_h^-/z} \sum_X \frac{1}{2 N_c} {\rm Tr } \biggr [  
    \langle  0\vert \gamma^+ \psi (0)  \vert h(P_h)X \rangle\langle X h(P_h) \vert \bar \psi(yl) \vert 0 \rangle \biggr ], 
\end{equation}                       
where we have suppressed gauge links. In the definition, the hadron moves in the $-z$-direction. In SIDIS $W^{\mu\nu}$ does not receive contributions involving 
matrix elements defined with chirality-odd operators. This is because the chirality-odd quark-gluon fragmentation function 
$\hat T^{(\sigma)}_F (z_1,z_2)$, which corresponds  to the chirality-odd quark-gluon distribution $T^{(\sigma)}_F(x_1,x_2)$ defined in Eq.(\ref{COTF}), is zero at $z_1=z_2$\cite{MeMe}. Beyond the tree level 
$W^{\mu\nu}$ can have chirality-odd contributions.   
\par 
At first look, the spin-dependent part of $W^{\mu\nu}$ in Eq.(\ref{WSIDIS}) has also two terms corresponding to the chirality-even 
contribution in Eq.(\ref{FW},\ref{TOW}) for Drell-Yan process. But, one can express Eq.(\ref{WSIDIS}) with the metric $g^{\mu\nu}$ of the transverse space defined with $P$ and $q$. The quantities in the two transverse 
spaces are given by: 
\begin{eqnarray} 
g_\perp^{\mu\nu} &=& g^{\mu\nu} -\frac{1}{P\cdot q} \biggr ( P^\mu (x_B P+ q)^\nu +P^\nu (x_B P+ q)^\mu \biggr),  \quad  P_{h\perp}^{\mu\nu} = g_\perp^{\mu\nu} P_{h\nu} = - z_h q_\perp^\mu + 
{\mathcal O}(P^2_{h\perp} ),
\nonumber\\
 g_\perp^{\mu\nu} &=& g_{\perp h}^{\mu\nu} + \frac{1}{z_h P\cdot q}\biggr ( P^\mu P_{h\perp}^\nu + P^\nu P_{h\perp}^\mu \biggr ) + {\mathcal O}(P^2_{h\perp} ).   \quad \tilde s_\perp^\mu = \frac{1}{P\cdot q} 
  \epsilon^{\alpha\beta\mu\nu} P_\alpha q_\beta s_\nu=\tilde s_{\perp h}^\mu.  
\end{eqnarray}
Here we assume that the spin vector $s$ in Eq.(\ref{SIDIS}) is in the transverse space defined with $P$ and $q$.  
Using these relations one can re-write $W^{\mu\nu}$ in Eq.(\ref{WSIDIS}) as:  
\begin{equation} 
 W^{\mu\nu} = - z_h^2 d(z_h) T_F (x_B,x_B) g_{\perp}^{\mu\nu} \tilde s_{\perp}^\rho 
   \frac{\partial \delta^2 (P_{\perp h})}{\partial P_{h\perp}^\rho}  +{\mathcal O}(\alpha_s).  
\label{HTC0}   
\end{equation}
This result is explicitly $U(1)$-gauge invariant. In comparison with $W^{\mu\nu}$ of Drell-Yan processes,
$W^{\mu\nu}$ of SIDIS has only one term which is spin-dependent. Therefore, one can only construct one weighted 
SSA weighted with $P_{h\perp }\cdot \tilde s_\perp$ in the integration over $P_{h\perp}$. The result is: 
\begin{eqnarray} 
\frac{ d\sigma \langle P_{h\perp}\cdot \tilde s_\perp \rangle } {d x_B d y d z_h} = \frac{\pi \alpha^2 e_q^2}{Q^2} \frac{1 +(1-y)^2}{y} 
   z_h d(z_h) T_F(x_B,x_B) + {\mathcal O}(\alpha_s).  
\label{WSSA}             
\end{eqnarray}    
This result agrees with that given in $\cite{BoMu}$. In \cite{BoMu} one uses Transverse-Momentum-Dependent(TMD) 
factorization and the weighted SSA is expressed with the transverse momentum moment of Sivers function. The moment is related to $T_F(x,x)$ at tree-level. 
\par 
We have shown that how QCD gauge invariance for each term of $W^{\mu\nu}$ in Eq.(\ref{FW}) for Drell-Yan processes is made. For the term with the derivative of $\delta^2(q_\perp)$ we have to sum all exchanges of $G^+$-gluons into gauge links as shown in Eq.(\ref{1b+1c}). This fact gives us the opportunity to simplify 
the calculation  of corrections from higher order of $\alpha_s$. 
It is noted that the derivative of the $\delta$-function for the momentum conservation comes from the expansion of transverse momenta of partons. In calculating  the twist-3 contribution with the derivative of the $\delta$-function, one can already neglect the transverse momenta of partons except those in the $\delta$-function. In this case, one essentially 
deals with the amplitude of the annihilation of an on-shell quark-antiquark pair combined with $G^+$-gluons from the lower bubble in Fig.1 into the virtual photon. Now the summation 
of all exchanges of $G^+$-gluons can be easily performed.     
After the summation the contribution with the derivative only comes 
from Fig.1a but with the bubbles defined with insertion of gauge links. Because of this, 
the virtual correction for the term with the derivative of $\delta^2(q_\perp)$ is completely determined by the quark form factor with the subtraction defined with the gauge links. The situation is similar to that of corrections beyond tree-level in TMD factorization discussed in \cite{TMDDY}. We note here that in the subtraction with the gauge links the collinear 
divergences will be subtracted.  The divergences from soft gluons are expected to be cancelled 
in twist-3 factorization\cite{QiuStS}.   
\par
It is noted that some weighted SSA observables do not receive contributions from the part of $W^{\mu\nu}$ with  $\delta^2(q_\perp)$. They receive contributions from the part with the derivative of $\delta^2(q_\perp)$ and the part with $q_\perp\neq 0$. 
From the above discussion, the virtual correction of these weighted SSA observables is determined by the quark form factor.
This will simplify the calculation of the virtual correction, because the virtual correction from exchanges of gluons in the left 
part of Fig.1b is more complicated than that from exchanges of gluons in the right part of Fig.1b. But, for the part with the derivative of $\delta^2(q_\perp)$ the two corrections are same. 
The same conclusion can also be made for SIDIS. 
The one-loop calculation of the virtual correction involving $T_F$  
for Drell-Yan processe in \cite{VoYu} and  for SIDIS in \cite{KVX} verifies our conclusion explicitly. 
\par 
To summarize: We have studied collinear factorization at twist-3 of the hadronic tensor $W^{\mu\nu}$ relevant for SSA in 
Drell-Yan processes with the emphasis of how to obtain results respecting gauge invariance of QCD, 
and with the hope to solve the discrepancy of SSA in the lepton angular distribution $A_N$. Our result for $A_N$ agrees our early result in \cite{MaZh}. It is noted that the result in \cite{MaZh} is derived in an approach 
different that that in this work.  
A by-product with our result of $W^{\mu\nu}$ is that  
we can construct another two observables of SSA in addition to $A_N$. Measuring these observables
will help to disentangle different contributions in $W^{\mu\nu}$. 
We also give 
the result of $W^{\mu\nu}$ relevant for SSA in SIDIS, where one can only obtain one SSA observable.  
We have shown that the QCD gauge invariance for different contributions in $W^{\mu\nu}$ is made in 
different ways. In showing this, we obtain an interesting result that the virtual correction to one structure function of  
$W^{\mu\nu}$, hence to some weighted SSA observables, is completely determined by the quark form factor with subtractions. This will simplify 
higher order corrections.

\par\vskip20pt
\noindent
{\bf Acknowledgments}
\par
The work of J.P. Ma is supported by National Nature
Science Foundation of P.R. China(No.11275244). The support from CAS center for excellence in particle 
physics is acknowledged.

\par

\par

\end{document}